\documentstyle[twoside,fleqn,espcrc2]{article}

\newcommand{\AmS}{{\protect\the\textfont2
  A\kern-.1667em\lower.5ex\hbox{M}\kern-.125emS}}

\title{Efficient Hadronic Operators in Lattice Gauge Theory}

\author{ UKQCD Collaboration presented by 
P. Lacock\address{DAMTP, University of Liverpool, \\ 
        P.O. Box 169, Liverpool L69 3BX, Great Britain}}%
       
\begin{document}

\begin{abstract}

We propose the study of non-local gauge invariant operators
to obtain an uncontaminated ground state for hadrons. The 
efficiency of the operators is shown by looking at the wave
function of the first excited state, which has a node as a      
function of spatial extent of the operator.

Liverpool Preprint LTH-339 (Nov. 1994); hep-lat/9411013

\end{abstract}

\maketitle

\section{Introduction}

In quenched lattice gauge theory, hadronic states are created by acting
with light quark (or anti-quark) creation operators on  the vacuum. 
The ground state mass is then determined in the large (euclidean) time
limit. In practice, however, only  
a limited range of $t$ is available, so methods are needed to
create the  hadronic ground state efficiently. 

Since the main contamination of the ground state signal at larger $t$ values 
comes from the first excited state, one requires a 
hadronic creation operator which maximises the ground state relative 
to this first excited state. 

One clear motivation for a trial hadronic operator comes from 
considering heavy quark hadrons: 
for heavy mesons ($c\overline{c}$ and $b\overline{b}$) the 
adiabatic approximation is well justified,
so that they can be modelled as heavy point particles (quarks) bound 
by a central potential $V(R)$ 
between static colour sources.
On a lattice efficient gluonic operators which
create such a colour flux between static sources
can be constructed using an iterative fuzzing
algorithm~\cite{smear}.
This prescription  creates
gluonic fluxes with a dominant ground state and very few excited
states.
To extend this approach to lighter quarks,
we use a lattice construction of a colour flux tube of
length $R$ to join two light quarks in a gauge invariant manner. 
By varying $R$ we can explore the relative amplitude of
ground state and
excited state hadron created.

\begin{figure}[htb]
\vspace{6cm}
\includegraphics{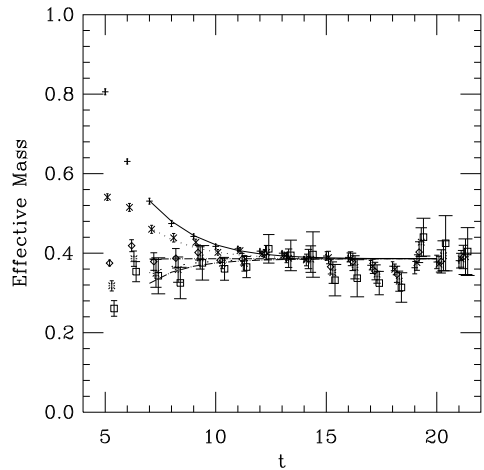}
\caption{The effective mass for the $\rho$ (in lattice units) using
the effective operators LL (+), and LF with R=4 ($\times$),
8 ($\diamond$), 10 ($\ast$) and 12 ($\Box$).}
\label{fig:largenenough}
\end{figure}

This relative amplitude is usually called the Bethe Saltpeter (BS)
wave function of the hadron. It is the overlap between  a quark and 
antiquark at distance $R$ apart and the hadronic state which is an 
eigenstate of the hamiltonian (transfer matrix on a lattice). We use 
a fuzzed gluon flux prescription introduced in an earlier work
with a similar 
approach~\cite{bs-gup} to join
the quarks, which corresponds to 
the adiabatic wavefunction as defined by~\cite{tn}.  Some previous 
work has used quark and antiquarks in the (spatial) coulomb gauge 
instead~\cite{bs-gup,tn}. 
This is less efficient (in our sense)
and also leads to problems with 
image sources in the spatial periodic boundary conditions.

\section{Lattice Measurements}

We use light quark propagators in the $24^3 \times 48$ configurations 
at $\beta=6.2$ and $K=0.14144$ using the clover action
obtained by UKQCD~\cite{ukqcd}. Our 
most comprehensive wave function results come from  an analysis of 12
configurations, although some quantities are available  from a larger
sample of 60 configurations. At these values of $\beta$ and $K$
the inverse lattice spacing is 
$a^{-1}=2.73(5) $GeV (determined from the string tension),
while $ m_{\pi} / m_{\rho} \approx 0.77$.  

\begin{figure}[t]
\vspace{6cm}
\includegraphics{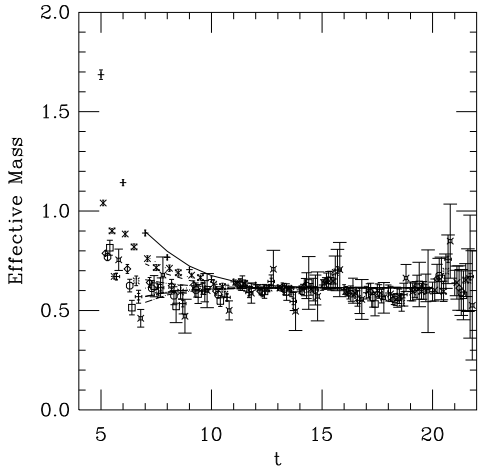}
\caption{The effective mass for the Nucleon. Here
we show the LLL, LLF and LFF operators for the various
$R$ values (see text).}
\label{fig:largenenough}
\end{figure}

We construct fuzzed gluon flux tubes
following the successful methods 
used for studying the potential~\cite{smear}:  
$$
 U_{new} = P_{SU(3)} (c U_{old} + \sum_1^4 U_{u-bends} ).
$$

Varying the parameters of the 
fuzzing  prescription in our case gives relatively little change. 
Hence we select a  smaller fuzzing level (5) 
and coarser fuzzing ($c=2$)
to minimise computer resources.

Since we have access only to light quark propagators from a single 
source ({\bf 0},0), we use a conventional local hadronic operator  at
the source. At the sink at time $t$, we use the spatially extended 
operator of length $R$ along a lattice axis.
A sum over spatial
positions (to have momentum zero) and a  sum over all 6 orientations (to
get correct $J^{PC}$) is used. 

Here we follow the definitions and construction of the hadronic
interpolating fields used by e.g. UKQCD 
~\cite{ukqcd}. 
The resulting hadronic correlation averaged over
all configurations is then fitted 
to a two exponential function
$$
<\! h(0)H^{\dag}(t,R) \!> \,= c_0(R) e^{-m_0 t} + c_1(R) e^{-m_1 t}
$$ 
We use a simultaneous fit to data at all $R$ values and $t$ values 
by making use of modified correlated $\chi^2$ fits, which 
models the correlations between different data points ~\cite{cor}.

In the following we use $R=4,8,10$ and 12, while
the local operator corresponds to $R = 0$.
It will be shown below that this range of $R$ values
is broad enough  
for the hadrons considered here. 
Statistical errors are determined using a bootstrap 
analysis.

For the mesons we form local-fuzzed (LF) correlations by 
replacing one local propagator in the usual (LL) formalism
by the fuzzed propagator discussed above.  
The effective mass results for 
vector meson are shown in Fig.~1.
There are five observables corresponding to LL, and LF 
with the four different $R$ values stated above. 

For the nucleon, we consider two different non-local operators.
These involve a di-quark separated from a quark (single
fuzzed - LLF) and an arrangement of three quarks all separated
(double fuzzed - LFF).  
Fig.~2 shows the results thus obtained for the nucleon. 

Both figures clearly show that, 
by using fuzzed non-local operators, the plateau in the data 
starts at smaller values of $t$ than for
the purely local observable.
The fuzzed data have larger errors at lower $t$ (compared to the
unfuzzed ones), but for larger  
$t$ the errors are comparable. The unfuzzed contribution thus has 
an accurate but irrelevant component.  

The two exponential fit, which is stabilised by
using several hadronic correlations      
to fit simultaneously,  
has the advantage that the  
ground state can be better isolated than 
by using a single exponential fit.

\begin{figure}[htb]
\vspace{5cm}
\includegraphics{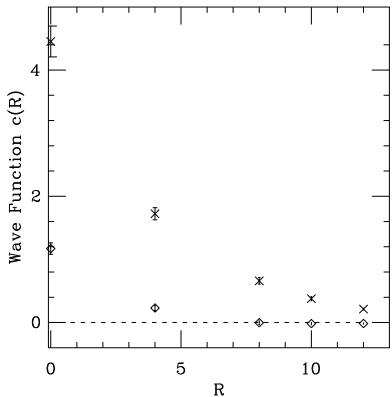}
\caption{The $\pi$ wave function for the ground state ($\times$)
and first excited state ($\diamond$), as functions of the length
of the gluonic flux tube.} 
\label{fig:largenenough}
\end{figure}

\section{Wave Functions}

To obtain a reasonably accurate determination of the wave function
of the first excited state
with our limited statistics, we fix the difference  
between the ground state and first excited state masses for
the hadrons from results obtained recently by
the UKQCD Collaboration ~\cite{ukqcdnew}.
These consist of hadronic propagators smeared       
at the source and sink (SS) and at the source only (SL) 
by applying the
Jacobi smearing method on the 
existing 60 configurations at the origin, obtained on the
$24^3 \times 48$ lattice at
$\beta = 6.2$ with $K=0.14144$ ~\cite{ukqcd}.   
A big advantage of having SS, SL and LL operators 
is that it allows a factorising fit.
These in turn provide tight 
constraints on the ground and excited state masses ~\cite{ukqcdnew}. 
We then proceed as before, fitting the local and non-local quark 
propagators calculated from the subset of 12 configurations to the
fit function given above.

\begin{figure}[htb]
\vspace{5cm}
\includegraphics{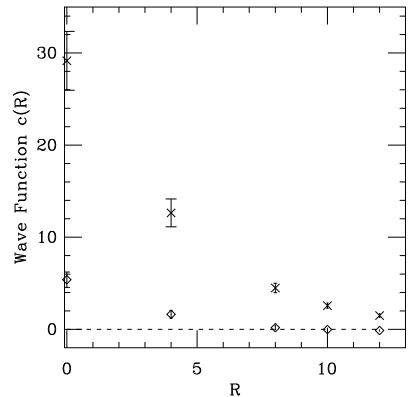}
\caption{As in Fig.~3, but for the Nucleon, using the LLF
effective operators.}
\label{fig:largenenough}
\end{figure}

The wave functions for the ground and first
excited states obtained from  
the local operators and those involving only one fuzzed link,
normalised at distance zero, are shown
in Figs.~3 and 4 for the $\pi$ and Nucleon
respectively. The wave functions for the $\rho$ and $\Delta$ show 
similar behaviour. The results for the ground state wave function 
are in agreement with those obtained previously 
 in the literature using similar  
gauge invariant definitions of the wave function ~\cite{bs-gup,tn}.

As far as we are aware, the excited state wave function has not been    
studied either for mesons or baryons.
The interesting feature that can clearly be seen for all the   
hadronic observables considered here is the presence of a node in the        
excited state wave function as a function of the length of fuzzed links 
connecting the propagator ends at the sink,     
where the ratio of ground state wf to excited
state wf becomes zero. This particular spatial
extent ($R \approx 8$ or $ \approx ~3  {\rm GeV^{-1}}$ in physical
units), which is
more or less the same for all the hadronic observables,
thus seems to be an optimal choice for producing a clean
ground state mass, since the  
contamination of the ground state by higher excited states 
has been minimised.


\begin{thebibliography} {99}


\bibitem{smear} S. J. Perantonis, A. Huntley and C. Michael,
Nucl.\ Phys.\ B326 (1989) 544.


\bibitem{bs-gup} R. Gupta, D. Daniel and J. Grandy,
Phys. Rev. D48 (1994) 3330. 

\bibitem{tn} K. B. Teo and J. W. Negele, Nucl. Phys. B (Proc. Suppl.)
34 (1994) 390.

\bibitem{ukqcd} UKQCD Collaboration, C. R. Allton et al., Phys Rev  
D49 (1994) 474.


\bibitem{cor} C. Michael, Phys. Rev. D49 (1994) 2616; \break 
C. Michael and A. McKerrell (in preparation)

\bibitem{ukqcdnew} UKQCD Collaboration, work in progress.


\end{thebibliography}
\end{document}